\documentclass[twocolumn,showpacs,showkeys,prc,aps]{revtex4}
\usepackage{graphicx}
  
\newcommand{ \be }{\begin{equation}}
\newcommand{ \ee }{\end{equation}}
\newcommand{ \bea }{\begin{eqnarray}}
\newcommand{ \eea }{\end{eqnarray}}
\newcommand{ \la }{\langle}
\newcommand{ \ra }{\rangle}

\newcommand{ \mpt } {{\la p_t \ra }}
\newcommand{ \cptt } {\kappa_{2,p_t} }

\newcommand{ \cptf } {\kappa_{4,p_t} }
\newcommand{ \dpt } {{\delta p_t }}
\newcommand{ \smpt } { \sigma_\mpt }
\newcommand{ \impt } {{\overline {p_t} }}
\newcommand{ \ispt } { \sigma_{p_t,inclusive} }

\newcommand{ \sd} {\sigma_{\mpt ,dynam} }
\newcommand{ \sigmastat } {\sigma_{\mpt ,stat} }
\newcommand{ \phisubpt } {\Phi_{p_t} }

\begin{document}

\title {
Mean $p_t$ fluctuations from 2- and 4-particle correlations
}


\author{Sergei A. Voloshin}
\affiliation{Wayne State University, 666 W. Hancock, Detroit,
Michigan}

\begin{abstract}
Recent results on event-by-event mean transverse momentum, $\mpt$, 
fluctuations in ultra-relativistic heavy ion collisions
are briefly reviewed. 
We conclude that the observed fluctuations are
in a rough agreement with that expected for the independent 
superposition of nucleon-nucleon collisions.   
We further discuss the possibility of the use of the forth order 
cumulants of the particle
transverse momentum distribution
in order to access the fluctuations related to collective phenomena.
\end{abstract}

\date{\today}

\maketitle


One of the most interesting
questions in the field of the ultra-relativistic heavy ion collisions 
is the question of the hadronization of the system. 
Is the system thermalized/equilibrated? 
Does the system evolution include the phase transition? 
Event-by-event fluctuations, e.g. of the mean transverse momentum, 
are considered to be one of the important tools 
to answer these questions~\cite{hh}.
The fluctuations depend on the nature of the phase transition. 
A first order phase transition may lead to 
large fluctuations in energy density
due to formation of QGP droplets{\cite{vanHove,kapusta}}. 
Second order phase transitions may lead to divergence in the specific heat; 
it would also increase the fluctuations in energy density due to
long range correlations in the system~\cite{stephanov98}.
One could observe them as fluctuations in mean transverse momentum
if matter freezes out at the critical temperature
$T_c$~\cite{stodolsky,shuryak98,stephanov98,berdnikov}.

The centrality dependence of the fluctuations is an
important observable. 
If the fluctuations are due to the particle production via 
some kind of clusters (e.g., resonances, strings,
(mini)jets, independent $NN$-collisions, etc.) and the relative
production of such clusters do not change with centrality,
the magnitude of the dynamical part in fluctuations
should be inversely proportional to the number of the clusters,
therefore to the particle multiplicity. 
New physics should appear as a deviation from such a dependence.   
There could be two reasons for the change in the centrality
dependence. 
First, such collective phenomena as
phase transition affect many particles in the system (unlike as
in the scenario of particle
production via a few particle clusters), and, therefore have different
multiplicity (centrality) dependence.
Second, new phenomena are expected to happen
at some critical particle density, which in turn depends on centrality. 

Consider fluctuations in the inverse slope of $m_t=\sqrt{m^2+p_t^2}$
distribution (effective temperature fluctuations).
Such fluctuations, for example, could be due to event-by-event 
fluctuations in radial expansion velocity. 
Then, depending on the slope, on average, the transverse momenta of 
all particles would be lower or  higher compared to the average over
all events. 
It results in (positive) correlations between transverse momenta 
of different particles~\cite{vkr,shuryak02}. 
One can quantify such correlations by the particle transverse momenta
covariance 
\be
\cptt \equiv {\rm cov} (p_{t,i}, p_{t,j}) 
=\la \dpt_i \dpt_j \ra_{i\ne j} = \sd^2,
\label{sd}
\ee
where $\dpt_i=p_{t,i}-\impt$, with $\impt$ being the inclusive mean
transverse momentum. 
It can be also useful to consider dimensionless quantities like
$\sd^2/\ispt^2$ or  $\sd^2/\impt^2$. 
The notation, $\sd^2$~\cite{vkr}, comes from
the fact that this quantity equals to the difference between
the actual variance of the event mean $p_t$  distribution
and the expected width due 
to the statistical fluctuations\footnote{
The statistical fluctuations are the ones one would observe 
in the case of independent particle production provided 
that the single particle
inclusive quantities remain the same as in real events. 
} in $\mpt$. The latter is due to the finite event multiplicity.   
\bea
\sigma^2_{\mpt}
&=& \frac{1}{M^2} ( \sum_i \dpt_i)^2 
\nonumber
\\
&\approx &\sigmastat^2 +\frac{M-1}{M} \sd^2,
\eea
where
\be
\sigmastat^2 = \frac{\ispt^2}{M},
\label{mptfluc}
\ee 
and $M$ is the multiplicity.

Event-by-event dynamical fluctuations have also been analyzed by
several experiments using
the so called $\Phi_{p_t}$~\cite{phipt} measure
(the approximate relation to $\sd^2$ is taken from~\cite{vkr}): 
\bea
\Phi_{p_t} &\equiv& 
\sqrt{\la (\mpt- M\impt)^2 \ra/\la M \ra }-\ispt
\nonumber \\
&\approx&
\frac{\sd^2 \langle M\rangle}{2\sigma_{p_t,inclusive}}.
\label{phiptdef}
\eea
and very close to it the difference factor 
$\Delta \sigma_{p_t}$~\cite{star}: 
$\Delta \sigma_{p_t} \equiv (\sqrt{\la M \ra} \ispt
-\sigma_{p_t,inclusive}) \approx \Phi_{p_t}$.

Using the above relations to compute $\cptt$ whenever
$\phisubpt$ being reported, we arrive to the conclusion that in
{\em central} collisions of heavy nuclei such as gold and lead, relative
fluctuations in mean transverse momentum, $\sd/\impt$ is of the order of
1 -- 1.5\%.  
For the 6\% most central collisions STAR reports~\cite{star} 
the preliminary results of  $\sd/\impt =1.2 \pm 0.2$.
The PHENIX measurements~\cite{phenix} have larger
uncertainties, but if averaged over different centralities (assuming
$\phisubpt$ does not change with centrality), this measurements yield
for the central Au+Au collisions the value of $\sd/\impt =1.4 \pm 0.9$.  
In Pb+Pb and Pb+Au collisions 
at CERN SPS ($\sqrt{s_{NN}}=17$~GeV) $\Phi_{p_t}$ 
have been measured by NA49~\cite{na49plb} and 
NA45/CERES~\cite{ceres}. 
The published NA49 result~\cite{na49plb} $\Phi_{p_t}=0.6 \pm 1$~MeV/c 
was obtained for the rapidity region $4.0<y<5.5$ 
and 5\% most central collisions.
The observed fluctuations are extremely small, but it would be incorrect to
compare these numbers with RHIC measurements, since NA49 results were
obtained in the forward rapidity region.
The CERES collaboration at the SPS has measured the fluctuations in the
central rapidity region. They report $\Phi_{p_t}=7.8 \pm 0.9$~MeV/c  
in central Pb+Au collisions~\cite{ceres}.
As seen from Eq.~\ref{phiptdef}, 
$\Phi_{p_t}$ is directly proportional to 
the number of reconstructed tracks used for its calculation 
(subject to acceptance cuts and tracking efficiency), which complicates
the comparison of the results from different experiments.
For a rough comparison one can take into account the CERES
multiplicity $\sim$130, $\langle \mpt\rangle \approx$420~GeV/c and
$\sigma_{pt,\; inclusive} \approx 0.270$~GeV/c. 
Then $\phisubpt \approx 8$~MeV corresponds 
to  $\sd/\impt \approx 1.4\%$. 
The centrality dependence of $\mpt$ fluctuations, whenever studied,
is consistent with particle production via clusters picture within
about ~30\%.

The dynamical part of mean transverse
momentum fluctuations has been measured at the ISR~\cite{isrpp}
in $pp$ collisions.
This was done by analyzing the multiplicity dependence of $\smpt$
under the assumption that in $pp$ collisions the dynamical part
in $\mpt$ fluctuations does not depend on multiplicity.
It was observed that $\sd/\langle \mpt\rangle \approx 12\%$.
Rescaling of this quantity with (the square root of) 
the ratio of multiplicity densities in
$pp$ and $Au+Au$ collisions yields the fluctuations in $Au+Au$:
$\sd/\langle \mpt\rangle \approx 0.8\%$, about 50\% less than
that observed in $AA$ collisions. Rescaling the quantity 
with the number of
participants gives even better agreement.
 
The non-zero value of $\cptt$ could be due to different reasons:
including such as resonance decays or jet production,
which are not
``interesting'' from the point of view of the collective phenomena 
under search in nuclear collisions.  
In principle, selecting specific pairs for the averaging in Eq.~\ref{sd} 
one can suppress or enhance the contribution to $\mpt$ fluctuations 
of different origin.
For example, one can correlate:
(1) particles in two disjoint pseudo-rapidity regions. The `gap' between
the two regions eliminates effects such as quantum statistics
(Bose-Einstein and Fermi-Dirac) correlations  or Coulomb final 
state interactions.  
(2) Positive particles with negative particles,
which is expected to enhance the contribution from 
resonances.
All those tricks do not solve the problem completely. 
Below we discuss the possibility of the use of 
4-particle correlations (cumulants) 
in order to evaluate genuine collective phenomena contribution to
mean $p_t$ fluctuations:
\be
\cptf=\la \dpt_i \dpt_j \dpt_k \dpt_m \ra -3 \la \dpt_i \dpt_j\ra^2 
\label{kappa4}
\ee
with all $i$, $j$, $k$, and $m$ being different.
The relative contribution of the collective phenomena
and particle production via clusters to
the cumulant $\cptf$ is enhanced by a factor of $\sim M_{total}/M_{cluster}$ 
compared to the relative contribution to $\cptt$. 
The notation  $M_{cluster}$ is used
here for the average cluster decay multiplicity, and $M_{total}$ is
the total event multiplicity.

In~\cite{ina} cumulants have been proposed for the study of
multiplicity fluctuations.
Recently, four particle azimuthal correlations (cumulants)~\cite{olli}
 have been successfully
used by the STAR Collaboration in the analysis of the azimuthal 
correlations for the measurements of elliptic flow~\cite{starflow3}.  
With modern statistics the technique works well even for relatively 
small signals (two particle correlations of the order of $10^{-4}$ and
smaller. This is just the range of the correlation magnitude needed for
the 4-particle correlations mean $p_t$ analysis (recall that
$\cptt/\mpt^2 \sim (1-1.5\%)^2$. In many particle cumulant mean $p_t$
analysis one can use the generating functions similar to the ones used
in~\cite{olli,starflow3}. 

The contribution to $\cptf$ from the inverse slope fluctuations
to the first order is proportional to the corresponding
cumulant of the inverse slope distribution. 
For the case of of an equal mixture of event with two different slopes
$T_1$ and $T_2$, the corresponding cumulant 
$\kappa_{4,T}=-(\Delta T)^4/8$, ($\Delta T= T_1-T_2$). 
Note that for some specific inverse slope 
distributions the cumulant could be small (for a rectangular
distribution with width $\Delta T$ it 
is $\kappa_{4,T}=-(\Delta T)^4/120$, and for 
a Gaussian distribution it is zero).  The very small values of the
cumulants may complicate the analysis.
The use of additional weights in Eq.~\ref{kappa4} could somewhat
help in this case. 

In {\bf summary}, we observe that while clear correlations 
in particle transverse momenta 
has been measured  in heavy ion collisions at SPS and RHIC, 
no evidence that the correlations are due to collective phenomena
have been found so far. 
The centrality dependence of the observed correlations, as well as
comparison to $pp$ collisions at similar energies, is roughly 
consistent  with the picture of particle production via clusters.  
In order to disentangle the contribution to the mean $p_t$
fluctuations coming from collective phenomena we propose to use
many-particle correlations.

The discussion with  A.~Poskanzer, R.~Bellweid, S.~Gavin,  J.-Y.~Ollitrault,
 and  C.~Pruneau are acknowledged. This work was supported in part by 
U.S. DOE  Grant No. DE-FG02-92ER40713.


\begin{thebibliography}{17}

\bibitem{hh} 
        H. Heiselberg, Phys. Rept. {\bf 351} (2001) 161.

\bibitem{vanHove}
	L. van Hove, Z.~Phys., \textbf{C 21}, 93 (1984).

\bibitem{kapusta}
	J. Kapusta, and A. Vischer, Phys. Rev., C {\bf 52}, 2725 (1995).

\bibitem{stephanov98}
	M. Stephanov, K. Rajagopal, and E. Shuryak, Phys. Rev. Lett.,
  {\bf 81}, 4816 (1998); Phys. Rev., D {\bf 60}, 114028 (1999).

\bibitem{stodolsky}
	L. Stodolsky, Phys. Rev. Lett, \textbf{75}, 1044 (1995).

\bibitem{shuryak98}
	E. Shuryak, Phys. Lett., \textbf{B430}, 9 (1998).

\bibitem{berdnikov}
	B. Berdnikov and K. Rajagopal, Phys. Rev., D \textbf{61}, 105017
  (2000).

\bibitem{vkr}
	S.A. Voloshin, V. Koch, and H.G. Ritter, Phys. Rev. D \textbf{60},
  024901 (1999).

\bibitem{shuryak02}
	 E. Shuryak, hep-ph/0205031 (2002).

\bibitem{phipt}
	M. Gazdzicki and S. Mrowczynski, Z. Phys., \textbf{C54}, 127 (1992).

\bibitem{star} 
	J. Reid for the STAR Collaboration, 
     	 talk at "Quark Matter 2001" conference, Nucl. Phys. \textbf{A698}
	(2002) 611c; 
	S.A.~Voloshin for the STAR Collaboration, talk at INPC 2001, 
        nucl-ex/0109006.

\bibitem{phenix}
	K. Adcox et~al., PHENIX Collaboration, nucl-ex/0203015

\bibitem{na49plb}
	 H. Appelshuser, et~al., NA49 Collaboration, Phys. Lett., 
	\textbf{B459}, 679 (1999).

\bibitem{ceres}
	H. Appelshauser for the CERES Collaboration, 
talk at "Quark Matter 2001" conference, Nucl. Phys. \textbf{A698}
	(2002) 253c;  

\bibitem{isrpp}
	K. Braune K., et~al., Phys. Lett., \textbf{B123}, 467 (1983).

\bibitem{ina} 
	P. Carruthers and Ina Sarcevic, Phys. Rev. Lett. \textbf{63},
1562 (1989);  H. C. Eggers, P. Lipa,  P. Carruthers, and  B. Buschbeck,       
  Phys. Rev.  {\bf D48}, 2040 (1993).       


\bibitem{olli}
       N. Borghini, P. M. Dinh, and J.-Y. Ollitrault,         
  Phys. Rev.  {\bf C63}, 054906 (2001); Phys. Rev.  {\bf C64}, 054901 (2001). 

\bibitem{starflow3} 
	C. Adler., et~al., STAR Collaboration, nucl-ex/0206001 (2002).

\end{thebibliography}
\end{document}